\RequirePackage{fix-cm}
\documentclass[smallextended]{svjour3}       
\smartqed  
\usepackage{graphicx}
\begin{document}

\title{A precursor interpretation for the Crab supernova 1054 A.D. very early light curve
}

\titlerunning{Precursor interpretation of SN 1054 A.D. }        

\author{J.E. Horvath$^{1}$}


\institute{J.E. Horvath \\
          {foton@iag.usp.br}\\
              $^{1}$ Universidade de S\~ao Paulo, Department of Astronomy IAG-USP\\
              R. do Mat\~ao 1226, 05508-090, Cidade Universit\'aria, S\~ao Paulo SP, Brazil\\
              Tel.: +55-11-30912710\\
              Fax: +55-11-30912806\\
               }

\date{Received: date / Accepted: date}

\maketitle

\begin{abstract}
In spite that the Crab supernova in 1054 A.D. was studied over the years, it is still not clear what type
of event produced the explosion. The most detailed and reliable source of the observed light curve
is recorded in the Chinese and Japanese chronicles, and suggests a quick dimming after a very bright initial peak.
We shall show in this work that the Crab event can be well explained by a type of precursor, a phenomenon emerging from
large supernova sampling and quite expected from Stellar Evolution considerations of low-mass
progenitors. This very early bright transient is followed by the explosion itself, likely a low-luminosity supernova from ``small iron core'' type instead of an electron-capture event. The latter conclusion stems from recent simulation work, which predict that an electron-capture supernova would render the magnitude to be much brighter for $\sim 3$ months, hence visible during daytime, and would not match the Chinese records.

\keywords{supernovae: individual: SN 1054 A.D. \and Optical transients \and supernovae: general}
\end{abstract}

\section{Introduction}
\label{intro}
The supernova 1054 A.D., which originated the Crab nebula and its associated pulsar are one of the most widely known and studied events in Astronomy. The connection of the Crab nebula with the historical supernova observed in 1054 A.D., with records in Chinese, Japanese and Western sources provided the best example of Baade and Zwicky's (1934) suggestion of the formation of neutron stars in events generally seen in {\it core-collapse} supernovae. However, this paradigm led to a series of questions which arise because the observed supernova contain some puzzling features, not observed in any other case in History.

Among the latter, we can count the lack of clear evidence of the supernova remnant itself, precluding a more direct identification of the type of event, and the reconstructed early light curve (mainly from Chinese sources) which has been difficult to understand in general. On the theoretical side, since the work of Nomoto and collaborators (1982), the Crab has been seriously considered as a candidate to be a case of {\it electron-capture supernova} (ECSN), in which a super AGB star explodes after the capture of electrons in an $O-Mg-Ne$ degenerate core. The progenitors of these super-AGB stars are expected to have $8-10 \, M_{\odot}$ in the MS, although the exact values depend on metallicity (Doherty et al. 2017). A neutron star formation is expected in many cases from these events too. Alternatively, low-mass iron cores collapses (CCSNs) could lead to similar explosions and leave a neutron star as observed, the latter featuring masses {\it below} the ``fixed'' value $\sim \, 1.25 \, M_{\odot}$ attributed for the ECSN events for the compact remnant (Horvath et al. 2022).

It is now known that the lightcurves of explosive events have revealed a variety, and a database on the long-term behavior of the lightcurves is better known because of detailed observations and theoretical work along the last decades. Nevertheless, several so-called ``supernova impostors'', transients and events that release $\geq 10^{51} erg$ but are not actual supernovae have been recognized and studied over the years. Eta Carina is one of the well-known cases that could have been identified as a supernova, although it belongs to a different class of transients (Davidson and Humphreys 2012).

In the particular case of the SN 1054 A.D., a total energy of the Crab $\leq 7 \times 10^{49} \, erg$ as an upper limit (see Hester 2008) has been reported. The optical lightcurve would have been similar to the cases of SN2005cs, SN2016ov and SN2018zd (Spiro et al. 2014, Hiramatsu et al. 2021). A sudden large drop of the brightness around $\sim 120-130 \, d$ should have been occurred in the SN 1054 event according to model simulations, although there are no reports of that in the available records. The supernova should have decreased by additional $\sim 6-7 \, mag$ after vanishing during daytime, to become completely invisible, even at night, after $653 \, d$ (Table 1). Note that these features are quite independent of the exact nature of the event, either a core-collapse of a small iron core progenitor or an electron-capture event, and any model must comply with the temporal behavior.

In either of these scenarios, ECSN or CCSN, the features of the initial lightcurve reconstructed for the Crab explosion are quite difficult to obtain. A bright maximum ($M_{V} \geq -18$) and rapid decay were observed, while the ECSNs and CCSNs scenarios predict in turn low explosion energies. We shall argue below that the very early lightcurve can be associated to a {\it precursor} of the type that are being seen in many supernova surveys (Section 3), and that the subsequent behavior of the light curve is the one expected from a low-mass iron core CCSN, not of an ECSN type (Section 4).

\section{Historical records}
\label{sec:1}

Several Chinese records refer to the ``guest star'' of 1054 A.D.. A comparative discussion by Breen and McCarthy (1995) concluded that the most likely date for its appearance in the 4th July with a fast (but not precisely determined) rise time $\leq \, 1 \, d$. The Chief of the Astronomical Bureau at K'Ai-Feng writings provide accurate times for the appearance of the object, comparing its brightness to Venus, which has $m_{V} = - 4.47$, interpreted as its maximum brightness (it should be noted that in spite of being able to distinguish $\sim 0.1 \, mag$, ancient Chinese astronomers have not registered the brightness variation in time of Venus, exceeding $0.5 \, mag$, Filipovi\'c et al. 2022). After the classical papers of Lundmark (1921), Mayall and Oort (1942) and Shkilovsky (1968), among others, the common assumption is that, after applying the extinction correction $A_{V} = 1.6$ (Miller 1973), the peak absolute magnitude should have been $M_{V} = -18$ or higher. However, the Japanese chronicles compared the event with a dimmer Jupiter ($m_{V} = - 2.2$, Green and Stephenson 2003), although the proximity of the Sun in the sky and some uncertainty in the dating made these reports not as solid. Therefore, it is possible that the brightness of the object at peak was slightly lower than Venus. We shall adopt the accepted value $M_{V} = -18$ hereafter as a lower limit. The precise value is impossible to refine, but fortunately not crucial here.

The same Chinese chronicles (Breen and McCarthy 1995) provide another important clue to the early lightcurve, saying that
the guest star was no longer visible during the day after $23 \, d$. This can be quantified by observing that the $m_{V} = - 0.24$ Saturn is often visible during the day, and leads after
extinction correction to a value $M_{V} \, \sim \, -14$. This estimation in uncertain by about $0.5$ mag, but overall it is fair to state that the transient faded by $\Delta M_{V} \sim \, 4$ in about three weeks. This  temporal feature is reliable from the Chinese imperial astronomers records and a prime feature to be understood.

The last temporal benchmark is related to the {\it late} lightcurve, i.e. the disappearance of the guest star from the night sky. Again, it is difficult to pin down exactly the sky background brightness, but this ``disappearance'' has been quantified as $M_{V} \sim  -7$ after $653 \, d$ (Breen and McCarthy 1995). It is unfortunate that no intermediate values are known, since supernova often present interesting and revealing behavior at these times.

The Western chronicles are scarce, more vaguely described (to the point of being doubtful that they describe a ``guest star'' event at all), and not due to astronomers. Some have considered the image of the Holy Roman Emperor Henry III considered as a proof that the 1054 A.D. was seen in Italy even though no written account of this specific observation remains (Hoffman and Gudrun 2021). A Middle East observation performed by Ibn Butlan is also known (Brecher 1978) but it does not contain substantial information either. An argument to associate a Flemish writing addressing the day of the Pope Leo IX death would need to bring the explosion to a much earlier date (Guidoboni et al. 1994). Because the death of the Pope Leo IX and the Great Schism of the Church, declared by the Patriarch of Constantinople on July 16th 1054, a heavy ``contamination'' (in the sense that some phenomena were reported to reinforce the holiness of the Pope, but could be completely unreal) of the reports is believed to have taken place, quite unrelated to the 1054 A.D. event itself, or in any case difficult to disentangle from the former. Certainly there is no consensus for a much earlier date for the explosion of the Crab supernova, which would also create big problems to sustain very high luminosities over a period of $> \, 3 \, months$ (April-July 1054) if accepted, thus it has been considered untenable (Breen and McCarthy 1995).

\section{The very early SN 1054 A.D. light curve as a precursor}
\label{sec:2}

Precursors (i.e. transient events) preceding supernovae have been reported in the literature (see, for instance Ofek et al. 2014 for a study of a Type IIn supernova sample). The statistics is very incomplete since large surveys are only now reaching the point when an evaluation
becomes feasible. In the case of the Crab as a confirmed low-mass explosion, a super-AGB character of the progenitor would enhance the expectations for episodic mass ejections or super-Eddington winds.

Precursors can be originated in a couple of different ways in this context.
The first is the ejection of mass, whose kinetic energy is converted into luminosity, with an efficiency factor $\epsilon$. The simplest estimation for the ejected mass stems from energetic considerations and reads

\begin{equation}
    M_{ej} \approx \epsilon {{2 L_{p} {\delta t}} \over {v^{2}}} \, = \, 10^{-2} M_{\odot}
    \times {{\biggl( {\epsilon \over {0.1}} \biggr)}} {{\biggl( {Lp \over {10^{9} L_{\odot}}}\biggr)}}
    {{\biggl( {\delta t \over {23 \, d}} \biggr)}}
    {{\biggl( {1000 \, km s^{-1} \over {v}}\biggr)}}^{2}
\end{equation}

where $L_{p}$ is the precursor luminosity, $\delta t$ its duration and $v$ the velocity of the ejecta.

A second mechanism, studied by Shaviv (2000, 2001) and applied to novae and other problems, is a super-Eddington wind outflow accelerating the matter, yielding a simplified expression

\begin{equation}
    M_{ej,wind} \approx W {{L_{p} {\delta t}} \over {c_{s}  c}} \, =
    \, 10^{-2} M_{\odot} \times {{\biggl( {W \over {10}} \biggr)}} {{\biggl( {Lp \over {10^{9} L_{\odot}}}\biggr)}}
    {{\biggl( {\delta t \over {23 \, d}} \biggr)}}
    {{\biggl( {{60 km s^{-1}} \over {c_s}} \biggr)}}
\end{equation}

with $W \approx 5-10$ an empirical constant derived for each particular problem (Shaviv 2000, 2001), and the sound velocity $c_{s}$ has been scaled to its expected value at the base of the wind $\approx 60 \, km \, s^{-1}$.

Once we apply these expressions to reproduce the very early light curve of the Crab, we obtain in both cases $M_{ej} \simeq 10^{-2} \, M_{\odot}$, provided the velocity $v$ is high enough or the sound velocity $c_{s}$ is not too small. This mass is very small, and although it can produce a bright transient, and more importantly fast rising ($t_{rise} \leq 1 \, d$) and decay ($t_{decay} \sim \, weeks$) as required by observational constraints, its effect on the later shock breakout of the supernova itself is small. On the other hand, if the {\it whole} early lightcurve has to be sustained by circumstellar material (CSM), the mass must be much higher ($\sim 0.3 M_{\odot}$, Smith 2013), and an initial rapid decay of at least 4 magnitudes in just three weeks appears much more extreme.

\begin{table}[htp]
\caption{Variation in the absolute magnitude $\Delta M_{V}$ of the remnant with time (after Breen and McCarthy 1995)}
\label{tab:truncatedsampling}
\centering
\begin{tabular}{l c c  }
\hline
      $|\Delta M_V|$ & $\Delta t$ (days) &   \\
    \hline
         $\geq 4$  & 23 &  \\
         $\sim 6-7$  & 653 & \\
         $\sim 4$& $\gg 300$ &  \\
\hline
\end{tabular}
\end{table}

A possibly related phenomenon are the rapidly evolving optical transients reported in the last decade (Poznanski et al. 2010, Drout et al. 2014, Arcavi et al. 2016). These events form a growing group, generally characterized by fast rising times ($\sim$ days), absolute magnitudes comparable to Type Ia supernovae and decay within a month. They have been also associated to the effect of a shock wave interacting with CSM or energy injection by a magnetar birth, both models implying a stellar explosion as an original trigger. The super-Eddington scenario would also be a suitable model for these transients on general grounds (Shaviv 2000, Ofek et al. 2014), although a detailed discussion has not been presented for this specific case.

A case study of the optical transient KSN2015K recently reported by K2/Kepler (Rest et al. 2018) serves to exemplify the possible relation with the early lightcurve of the Crab supernova. This transient raised in $\sim 2 \, d$ and decayed on a $\sim \, 1 \, month$ timescale, with a peak luminosity a factor of $\sim \, 5$ higher that the reconstructed for the Crab. However, as stated the peak magnitude of the Crab could have been brighter that $-18$ when first spotted the morning of July 4th (Green and Stephenson 2003, Smith 2013), making the KSN2015K almost a perfect fit to the lightcurve without any modification. If a transient like this happened in the SN 1054 event, the time to decay to a magnitude that would have rendered the transient unobservable during daytime, as reported by the Chinese, come out automatically right (Fig. 1) and is consistent with the historical observations (Murdin and Murdin 1985, Breen and McCarthy 1995).

\begin{figure}
\centerline{\includegraphics[width=0.98\linewidth]{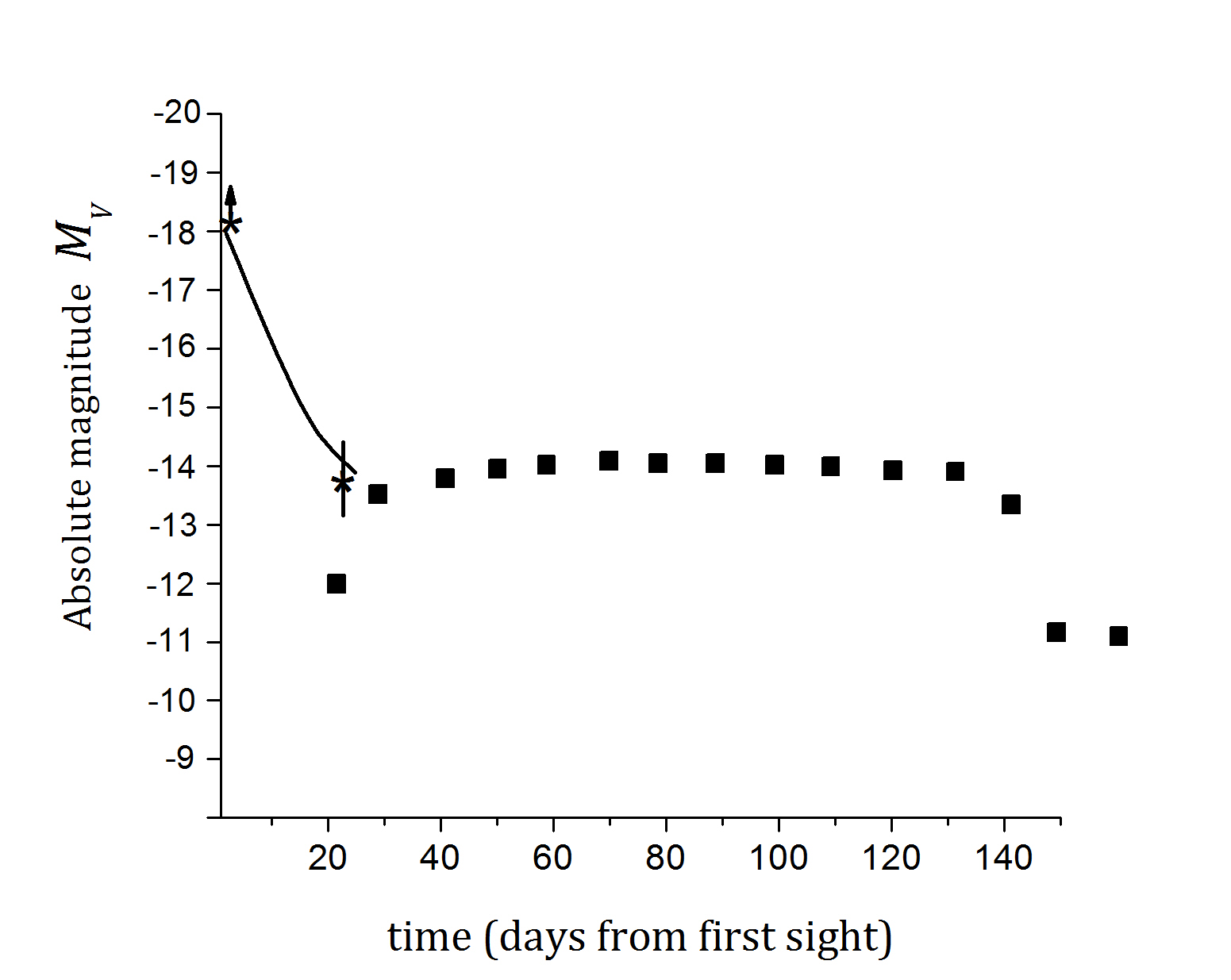}}
\caption{The very early ($t \leq 23$ d) and early ($t \geq 23$ d) light curves of the 1054 A.D. explosion. Asterisks are the reconstructed values according to the Chinese records. The solid curve corresponds to  the reported KSN2015K transient, with the maximum of the latter matched to the time of explosion. The squares are the $V$-band calculation of Kozyreva et al. (2021) corresponding to the s9.0 model, assumed to explode $\sim 3$ weeks after the precursor rise. According to the Chinese records, the latter curve should drop below $M_{V} = 7.5$ in April, 1056, 21 months after the first sight (Table 1).}
\label{fig:posterior}
\end{figure}

In the CSM breakout scenario, a rapid rise $ < 1$ day due to shock breakout diffusion $t_{d} \sim 30 R/c$ could be achieved if the CSM is standing at $\sim 10^{13} \, cm$. A compact progenitor of the carbon/oxygen type would have a much smaller radius. Only a small mass fraction $\leq 0.1 \, M_{\odot}$ would have to be ejected on a timescale $t_{csm} = R/v_{csm}$ over the last years before the explosion to explain the very early light curve (Rest et al. 2018).

A super-Eddington model will rise even faster and require a somewhat smaller ejected mass.
This model seems consistent with the evidence available the observed rise and decay in the KSN2015K too. The decay of the optical transient suggests that no other energy source like $^{56} Ni$ or continuous injection by a central magnetar (not expected for the ``normal'' magnetic field associated to the Crab pulsar (Allen and Horvath 2004, Kasen et al. 2016, Shukbold and Woosley 2016) is involved in the transient (Rest et al. 2018). To be sure, the growing group of transients include short events which are {\it not} consistent with shock breakout, and that can be interpreted within an ejected mass with radioactive decays as well (Ofek et al. 2021). Without any photometry and a scarce temporal information, we may never know which kind of precursor happened in SN 1054.

\section{The subsequent and long-term behavior of the lightcurve}
\label{sec:3}

An extensive study of low-luminosity explosions (Kozyreva et al. 2021) shows that it would be very difficult to understand the very early lightcurve from the theoretical point of view invoking just a ``bare'' explosion.It remains to be seen whether this feature stands with other calculations with varying physical content.

If the calculations of Kozyreva et al. (2021) are taken as a guide, we further conclude that the ECSN lightcurve based on the e8.8 model does not fit the Crab event, because it will render SN 1054 observable during the daytime for $\sim 4$ months. Explosions based on the more compact progenitors z9.6 and s9.0 will be acceptable, for a range of metallicities and explosion energies $few \times 10^{49} erg$. However, and guided by the simulations, this would mean that the Crab was {\it not} a ECSN, but rather a low-mass iron core CCSN. Moreover, this  conclusion is actually independent of the precursor interpretation of the very early light curve, and relies on the Chinese report on the supernova becoming invisible during the day after $23 \, d$. An ECSN would be very visible during the day because it would be about 2 magnitudes brighter (looking like Jupiter) according to the detailed models of Kozyreva et al. (2021). As a final corollary, the Crab pulsar mass is predicted to be below $1.25 M_{\odot}$ to achieve consistency within the low-mass iron core CCSN model.

In addition, it should be noted that the observed analogues SN2005cs and related events seem too dim at peak magnitude to be compared to the Crab case. However, the small overall mass $\sim 5 \, M_{\odot}$ in the Crab Nebula, the chemical evidence for a progenitor $\simeq 8 \, M_{\odot}$ and the low $^{56}Ni$ present (Smith 2013) remain as strong indications of a low-energy explosion, which also produced a neutron star. This is why we believe that a bright precursor interpretation, followed by the explosion itself, is a good fit to the whole picture.

\section{Conclusions}
\label{sec:4}

In summary, we have interpreted the initial phase of the supernova 1054 A.D. as a bright transient precursor, followed by the explosion itself after $\sim 3 \, weeks$. This interpretation has been motivated by the expected features of a low-mass progenitor near the explosion, which also predicts a plateau-type visual magnitude lasting 3-4 months but at a level that will not re-brighten the event (which was not reported). Bright optical transients such as KSN2015K possess all the features of the very early SN 1054 light curve and have been brought as comparison examples, although their actual relevance to the event remains to be proved.

In fact, it is quite remarkable that a lightcurve from a recent FELT event (KSN2015K) can fit accurately the reconstructed historical time behavior with just a slight (if any) brightness scale-down amounting to a small numerical factor. This can not, of course, taken as a proof, but certainly suggests a kinship of optical transients and SN precursors, a central thesis for our model.

Interestingly, the idea that CSM may be involved in supernovae with an early luminosity excess was also present in Goldberg and Bildsten (2016), Morozova et al. (2020) and Moriya et al. (2020). Smith (2013) has stressed the apparent incompatibility of the low-energy hypothesis of the Crab supernova with the high-luminosity early lightcurve and offered a detailed discussion of a CSM hypothesis. His view is different from our precursor transient interpretation that would explain the very early phase, while the well-known ordinary expansion of the supernova would have taken over after $\sim 3 \, weeks$ and be responsible for the long-term behavior, but this will be problematic if the calculations of Kozyreva et al. (2021) stand for the ECSN cases due to the non-visibility of SN 1054 A.D. supernova during daytime after $23 \, d$.

There are a few observational tests that may resolve the issue of the type of event. The most likely one would be the detection of ``light echoes'' of the event, which have the potential to reveal the spectrum and possibly its time evolution (Rest et al. 2008). However, in spite of the efforts over the years, the light echoes of the SN 1054 A.D. have not been detected yet. Therefore, other evidences such as the nucleosynthesis yields and structure of the remnant should be analyzed. We have remarked above the difficulties to study the latter. In fact, it is not clear that any other event of the ``SN 1054 A.D.-type'' has been observed, although this would not be so surprising because the derived number of rapidly evolving luminous transients is $\sim$ a few percent of the whole core-collapse SN (Drout et al. 2014). The follow-up of these fast transients could revel a weak SN after a few weeks, provided they are close enough and a strategy to identify and locate them quickly is developed (Inserra 2019). This would resemble the cases in which a long GRB is followed by a supernova, {\it mutatis mutandis}, and would constitute a crucial observation to assess the precursor scenario discussed above.

According to our picture, around $\sim \, 1 \, month$ the light curve would have leveled off due to the underlying supernova emergence, and decayed very slowly until $120-130 \, d$ (Fig. 1). Although the suggestions of a Type IIn-P from a small iron core explosion (Kozyreva et al. 2021) and ECSN (Nomoto et al. 1982) would be difficult to distinguish as low-energy events at late times in historical supernovae, the theoretical models of Kozyreva et al. (2021) disfavor the ECSN explosions as a model for the SN 1054 because they would be far too bright (by two magnitudes) to match the Chinese temporal records. Therefore, an agreement of present models and the SN 1054 reconstruction suggests a core-collapse SN, not an electron-capture one, independently of the origin of the very early bright lightcurve and related to the disappearance during the daytime only. Since the available models are scarce, and there are caveats which apply to a handful of points, we are not claiming anything definitive, but rather point out an alternative interpretation of the Crab event that could tie optical transients to the exploding stars, and tentatively identify which one produces the observed phenomenology. It may be that the confirmed diversity of supernova events and associated precursor/transients could be crucial to understand a millennium-old puzzle in a paradigmatic case.

\section*{Acknowledgements and declarations}

* The author wish to acknowledge the financial support of the Fapesp Agency (S\~ao Paulo) through the grant 2020/08518-2 and the CNPq (Federal Government, Brazil) for the award of a Research Fellowship.The members of the GARDEL Group at USP are acknowledged for their encouragement and scientific advise in these topics. An anonymous Referee helped to improve the first version with several remarks and
suggestions.

\bigskip
\noindent
* Informed consent does not apply to this work.

\bigskip
\noindent
* Data availability does not apply, no data beyond the one publicly available in the cited works has been employed.

\bigskip
\noindent
* The author declares no competing interests.

\bigskip
\noindent
* All the conception, execution and writing of the paper was performed by the author.



\vfill\eject

\end{document}